\documentclass{PoS}

\title{Simultaneous radio/X-ray observations of Cir X-1}

\ShortTitle{Simultaneous radio/X-ray observations of Cir X-1 \hspace{5.5cm} {\rm V. Tudose et al.}}

\author{V. Tudose,$^{1,2}$ P. Soleri,$^{1}$ R.P. Fender,$^{3,1}$ P.G. Jonker,$^{4,5,6}$ 
M. van der Klis,$^{1}$ A.K. Tzioumis,$^{7}$ R.E. Spencer$^{8}$ \\ 
\llap{$^1$}Astronomical Institute `Anton Pannekoek', University of Amsterdam \\
\llap{$^2$}Astronomical Institute of the Romanian Academy \\
\llap{$^3$}School of Physics and Astronomy, University of Southampton \\
\llap{$^4$}Netherlands Institute for Space Research \\
\llap{$^5$}Harvard-Smithsonian Center for Astrophysics \\
\llap{$^6$}Astronomical Institute, University of Utrecht \\
\llap{$^7$}Australia Telescope National Facility \\
\llap{$^8$}Jodrell Bank Observatory, University of Manchester \\

E-mail: \email{vtudose@science.uva.nl,psoleri@science.uva.nl}}

\abstract{We present a partial analysis of a multi-wavelength study of the X-ray binary Cir X-1, a system
harboring the most relativistic outflow in our galaxy so far. The data were taken (almost) 
simultaneously in radio and X ray during a survey carried out in October 2000 and December 2002. 
Cir X-1 was observed at the radio frequencies of 4.8 and 8.6 GHz with the Australia
Telescope Compact Array (ATCA). In the X-ray spectral domain we used the
Rossi X-Ray Timing Explorer (RXTE). We found strong evidence for flaring activity 
in radio not only at the periastron but also at the apoastron passages. A comparison of our data 
against different correlations between radio and X ray found in other neutron star systems shows 
that Cir X-1 does not seem to follow the general trend. However, the fact that Cir X-1 is an `exotic' 
X-ray binary makes any interpretation more complicated.} 

\FullConference{VI Microquasar Workshop: Microquasars and Beyond\\
		September 18-22 2006\\
		Societ\`a del Casino, Como, Italy}

\begin{document}

\section{Introduction}

Cir X-1 is a very unusual neutron star X-ray binary system with a changing 
behaviour mimicking both a Z and an Atoll source. It undergoes outbursts at 
X-ray, infrared and radio wavelengths with a period of 16.6 days. While 
the radio flares reached up to 1 Jy in the late 1970s, they have only been 
observed at the tens of mJy level ever since (e.g. \cite{Ste91}). A secular tendency toward 
lower emission rates in the 2-10 keV band is also evident in X rays in the last decade (e.g. \cite{Tud06}). 
The binary lies within an arcmin scale synchrotron nebula and seems to harbor the most 
relativistic outflow observed in our galaxy so far \cite{Fen04}. 

\section{Observations}

We observed simultaneously or quasi-simultaneously ($\sim$ hours delay) Cir X-1 
in radio and X rays for several days in 2000 October and 2002 December. The radio 
data, at 4.8 and 8.6 GHz, were acquired using the Australia Telescope Compact Array (ATCA). 
In the X-ray domain we made use of the Rossi X-Ray Timing Explorer (RXTE). Table 1 presents 
the epoch of observations together with the corresponding orbital phases as calculated 
using the radio ephemeris from \cite{Ste93}. 

\section{Results}

{\it Behavior of the radio flares.} The 16.6 days periodic outbursts of Cir X-1 are interpreted as 
enhanced accretion near the periastron passage. As expected, we observed an increase in the radio 
flux densities near the orbital phase $\phi=0.0$. But in addition, in our data set, both of the 
observations near phase $\phi=0.5$ (2000 October 09/10 and 25/26) show strong evidence for flaring 
activity of comparative magnitude as at $\phi=0.0$. The radio variability is at time-scales of hours 
and the spectrum changes dramatically during the flares, with a clear tendency towards flattening on 
the rising phase of the flare and steepening afterward. 

\begin{table}
\footnotesize
 \centering
\begin{center}
  \caption{Orbital phase of the start of the ATCA radio observations (determined with the ephemeris from 
\cite{Ste93}). The RXTE X-ray observations were simultaneous or almost simultaneous ($\sim$ hours 
delay) with the radio data.}
   \begin{tabular}{lc}
\hline
\hspace{1.0cm}   Epoch  &  \hspace{2cm} Phase $\phi$ \\
\hline
\hspace{0.3cm} 2000 Oct 01 & \hspace{2cm} 0.93 \\ 
\hspace{0.3cm} 2000 Oct 07/08 & \hspace{2cm} 0.36 \\ 
\hspace{0.3cm} 2000 Oct 09/10 & \hspace{2cm} 0.48 \\ 
\hspace{0.3cm} 2000 Oct 14/15 & \hspace{2cm} 0.78 \\ 
\hspace{0.3cm} 2000 Oct 19 & \hspace{2cm} 0.02 \\ 
\hspace{0.3cm} 2000 Oct 20/21 & \hspace{2cm} 0.14 \\ 
\hspace{0.3cm} 2000 Oct 23 & \hspace{2cm} 0.26 \\ 
\hspace{0.3cm} 2000 Oct 25/26 & \hspace{2cm} 0.45 \\ 
\hspace{0.3cm} 2002 Dec 02/03 & \hspace{2cm} 0.88 \\ 
\hspace{0.3cm} 2002 Dec 03/04 & \hspace{2cm} 0.94 \\ 
\hspace{0.3cm} 2002 Dec 04/05 & \hspace{2cm} 0.01 \\ 
\hspace{0.3cm} 2002 Dec 05/06 & \hspace{2cm} 0.07 \\ 
\hspace{0.3cm} 2002 Dec 06/07 & \hspace{2cm} 0.13 \\ 
\hspace{0.3cm} 2002 Dec 07/08 & \hspace{2cm} 0.19 \\ 
\hspace{0.3cm} 2002 Dec 08/09 & \hspace{2cm} 0.25 \\ 
\hline
\end{tabular}
\end{center}
\end{table}

\begin{center}
\begin{figure}
\includegraphics*[width=1.0\textwidth]{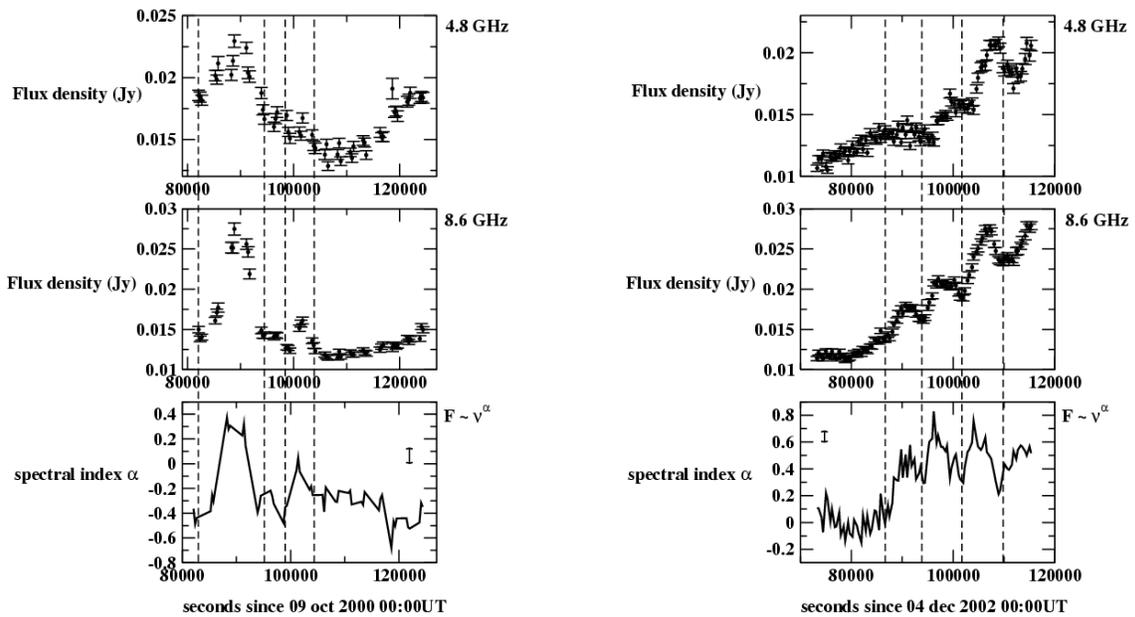}
\caption{ATCA 5-min averaged radio light curves at 4.8 and 8.6 GHz on 2000 October 09/10 
($\phi$=0.48) and 2002 December 04/05 ($\phi$=0.01). The spectrum between 4.8 and 8.6 GHz 
($F_{\nu} \propto \nu^{\alpha}$) was determined from unaveraged data. The typical errors 
for the spectral index plots are indicated separately. All the errors are at 1$\sigma$ level. 
The vertical dotted lines isolate the individual flares.}
\end{figure}
\end{center}

{\it Radio and X-ray luminosities.} In \cite{Mig06} the authors studied the correlation between radio 
and X-ray luminosities of neutron star X-ray binaries. In Fig.2 we added the data on Cir X-1 to 
their sample. The radio luminosities ($\propto \nu F_{\nu}$) were determined at 8.6 GHz from the ATCA data set, 
while the X-ray 
luminosities were calculated using the RXTE/ASM measurements in the 2-10 keV band. During 2002 
December, Cir X-1 was weaker in radio than in 2000 October, with luminosities 
$\leq 10^{30}$ erg/s (except for the two epochs near $\phi$=0.0, December 04/05 and December 05/06). 
In 2000 October, the luminosities were always higher than $10^{30}$ erg/s. \\

\begin{figure}
\begin{center}
\includegraphics*[width=0.55\textwidth]{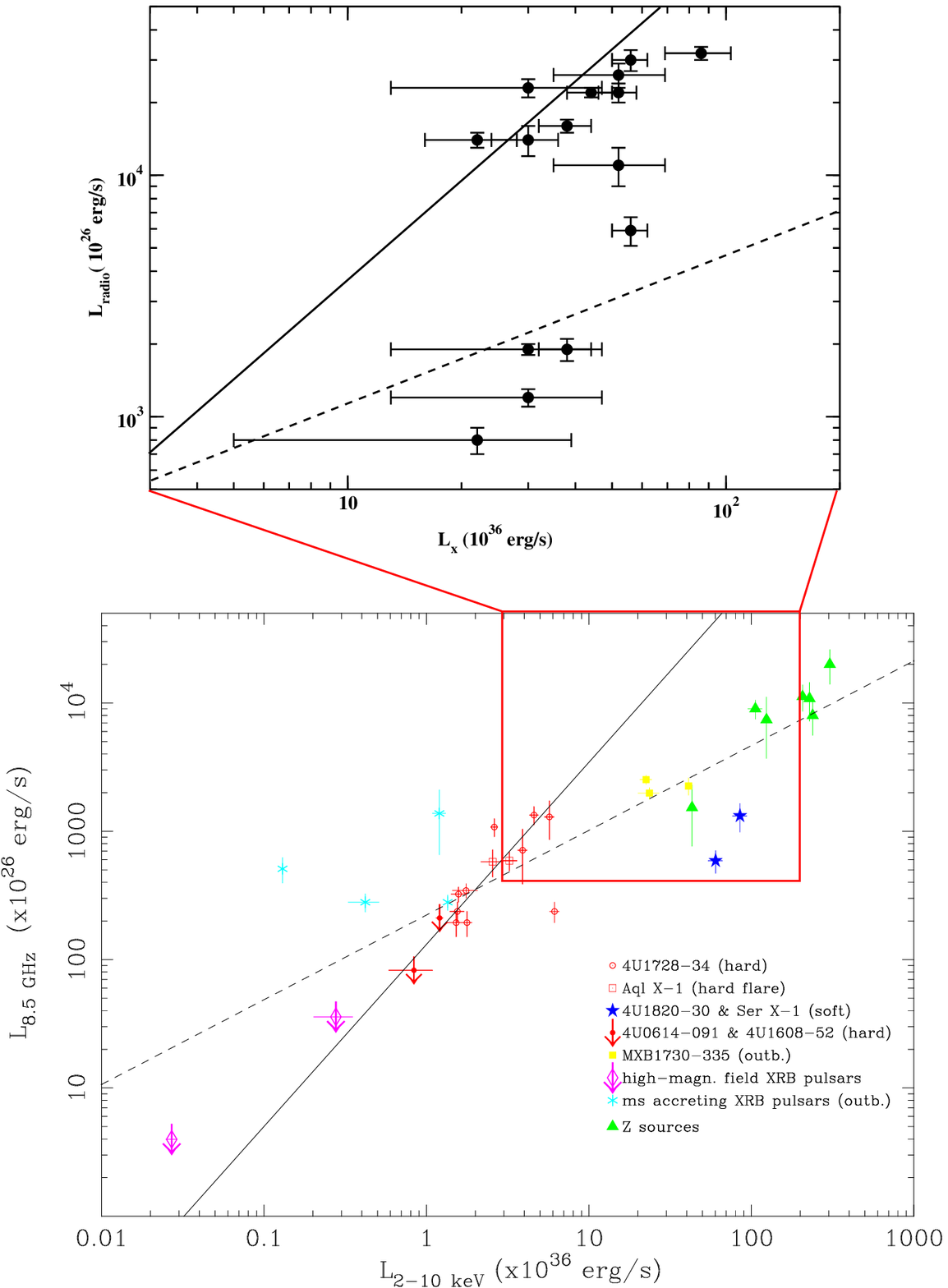}
\caption{The radio luminosity (ATCA data at 8.6 GHz) vs X-ray luminosity 
(RXTE/ASM data in 2-10 keV band) for Cir X-1 compared to the correlation 
in \cite{Mig06} for neutron stars in hard state (solid line) 
and all the Atolls and Z sources in their sample(dashed line). 
The errors in our plot are at 1$\sigma$ level.}
\end{center}
\end{figure}

{\it Radio luminosity and X-ray timing.} We created power density spectra (PDSs) from the X-ray data 
from segments 
of 128 s using fast Fourier transform techniques (e.g.\cite{Kli89}), averaged them together, 
subtracted the background, and fitted them with a multi-Lorentzian model. 
The frequencies quoted throughout are the frequencies at which the Lorentzian contributes 
most of its power: $\nu_{max}=\sqrt{\nu_0^{2}+HWHM^2}$, where $\nu_0$ is the centroid frequency of the 
Lorentzian (e.g. \cite{Bel02}). The spectra from 2002 December 02/03 and 03/04 were featureless and 
therefore not accounted 
for in Fig.3 (the appearance of PDSs of Cir X-1 depends on the orbital phase and ``position'' in the 
hardness-intensity diagram; e.g. \cite{Shi98}). In identifying the quasi-periodic 
oscillations (QPOs) in the PDSs, for practical reasons, we adopt the terminology for Atoll sources 
(e.g. \cite{Bel02}). Namely, we designate the QPOs in the 1-50 kHz range as the low-frequency QPOs, 
$L_{h}$. Previously, it was observed that the 
corresponding frequency $\nu_{h}$ shifts towards higher values with increasing orbital phase 
(e.g. \cite{Shi96}). PDSs averaged on data partitioned in blocks of a few hours near $\phi$=0.0 and 
$\phi$=0.5 (i.e.2000 October 09/10, 19, 20/21 and 25/26), show 
significant shifts in the $\nu_{h}$ and variations of rms $L_{h}$ at timescales of hours (the points 
are embedded in ellipses in Fig.3). Similar behaviour is observed for other orbital phases, but the changes are 
of a smaller magnitude. Whether this is real or an artifact is under investigation. We further show in Fig.3 
the correlations between the 
luminosity in radio and rms $L_{h}$ and $\nu_{h}$ found in \cite{Mig05} in ``well-behaved'' 
neutron star X-ray binaries. Clearly, Cir X-1 does not follow the trend. 

\begin{figure}
\begin{center}
\includegraphics*[width=0.9\textwidth]{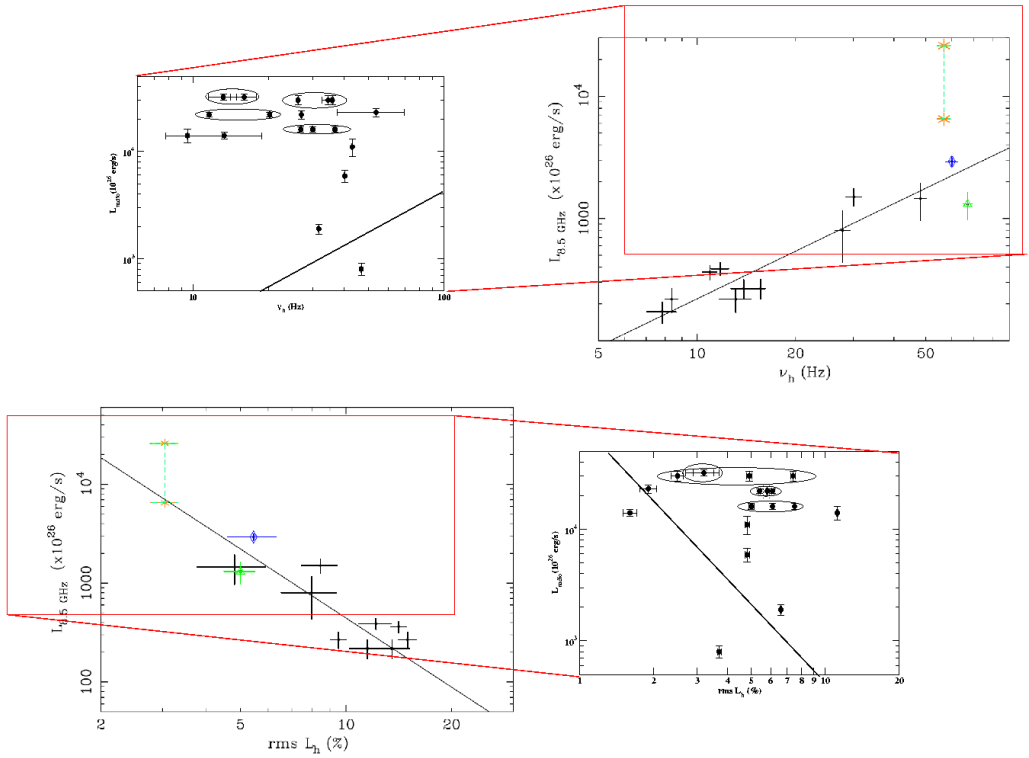}
\caption{Correlations for the radio luminosity and $\nu_{h}$ and rms $L_{h}$ from \cite{Mig05} for the 
neutron star X-ray binaries 4U 1728-30 (dots), Ser X-1 (open triangle), 
MXB 1730-335 (open diamond) plus the "peculiar" Atoll GX 13+1 (the asterisks 
represent the range of values in outburst), and our corresponding data for Cir X-1. 
In our plots, the points inside the ellipses represent values determined by partitioning 
the X-ray data close to $\phi$=0.0 and $\phi$=0.5 in blocks of a few hours each.}
\end{center}
\end{figure}

One way to explain the `anomalous' behaviour of Cir X-1 presented in Figs. 2 and 3 could be to suggest that the 
source is simply too radio bright, which is compatible with the emission being highly beamed. 

\acknowledgments

The Australia Telescope is funded by the Commonwealth of Australia for operation as a 
national facility managed by CSIRO.

\end{document}